\documentclass[12pt,english,floatfix,superscriptaddress,aps,preprint,amsmath,amssymb]{revtex4}
\usepackage[latin9]{inputenc}
\usepackage{amsmath}
\usepackage{amssymb} 
\usepackage{amsbsy}
\usepackage{amsfonts}
\usepackage{amsopn}
\usepackage{amstext}
\usepackage{graphicx}
\usepackage{esint}
\usepackage{hyperref}

\usepackage{amssymb}
\usepackage{amsfonts}
\usepackage{amsmath}
\usepackage{graphicx}
\usepackage[english]{babel}
\usepackage{color}
\usepackage[dvips]{epsfig}
\usepackage[dvips]{graphicx}
\usepackage{float}
\usepackage{units}
\usepackage{textcomp}
\usepackage{babel}

\begin{document}

\title{Dirac equation in very special relativity for hydrogen atom}

\author{R. V. Maluf}
\email{r.v.maluf@fisica.ufc.br}
\affiliation{Universidade Federal do Cear\'a (UFC), Departamento de F\'{\i}sica, Campus do Pici, Caixa Postal 6030, 60455-760, Fortaleza, Cear\'{a}, Brazil}

\author{J. E. G. Silva}
\email{euclides@fisica.ufc.br}
\affiliation{Universidade Federal do Cear\'a (UFC), Departamento de F\'{\i}sica, Campus do Pici, Caixa Postal 6030, 60455-760, Fortaleza, Cear\'{a}, Brazil}

\author{W. T. Cruz}
\email{wilamicruz@gmail.com}
\affiliation{Instituto Federal de Educa\c{c}\~{a}o, Ci\^encia e Tecnologia do Cear\'{a} (IFCE),
Campus Juazeiro do Norte, 63040-000, Juazeiro do Norte, Cear\'{a}, Brazil}

\author{C. A. S. Almeida}
\email{carlos@fisica.ufc.br}
\affiliation{Universidade Federal do Cear\'a (UFC), Departamento de F\'{\i}sica, Campus do Pici, Caixa Postal 6030, 60455-760, Fortaleza, Cear\'{a}, Brazil}

\date{\today}

\begin{abstract}
In this work, we study the modified Dirac equation in the framework of very special relativity (VSR). The low-energy regime is accessed and the nonrelativistic Hamiltonian is obtained. It turns out that this Hamiltonian is similar to that achieved from the Standard Model Extension (SME) via  coupling of the spinor field to a Lorentz-violating term, but new features arise inherited from the non-local character of the VSR. In addition, the implications of the VSR-modified Lorentz symmetry on the spectrum of a hydrogen atom are determined by calculating the first-order energy corrections in the context of standard quantum mechanics.  Among the results, we highlight that the modified Hamiltonian provides non-vanishing corrections which lift the degeneracy of the energy levels and allow us to find an upper bound upon the VSR-parameter.
\end{abstract}


\maketitle

\section{Introduction}

Despite the great success of the Standard Model of particle and of the General Relativity, both based on the Lorentz symmetry, some open issues as the non-vanishing neutrino mass \cite{Ma:2001mr} and the dark matter \cite{Rubim:1970,Goodman:1984dc} boosted the quest for physical theories with other symmetry groups.
In order to solve these and other problems of high-energy physics, some models propose enhancing the Lorentz algebra by modifying the dispersion relations and the dynamics of particles. The doubly special relativity (DSR) assumes the existence of two invariants, namely, the velocity of light and the Planck energy scale (or the Planck length) \cite{AmelinoCamelia:2000ge,Magueijo:2001cr}. In another approach the Standard Model Extension (SME) conjectured the existence of fixed background tensor fields. They couple with the standard model fields giving rise to the Lorentz
violating effects \cite{Colladay1997,Colladay1998}. The parameters representing this kind of violation are thought as the vacuum expectation values of some Lorentz fields belonging to the underlying theory. The experimental consequences have been scrutinized over the years and were still waiting for some sign of confirmation, see e.g. \cite{Kostelecky2011,Liberati} and references therein.

In a novel framework, Cohen and Glashow proposed a relativity theory that preserves the usual energy-momentum dispersion relation although it is not invariant under the full Lorentz group \cite{ref1,ref2}. Instead of using the Lorentz group $SO(1,3)$, Cohen-Glashow
realized that a proper subgroup of the Lorentz group can lead to the conservation laws and the well studied effects of the Special Relativity (SR). There are two subgroups fulfilling these requirements, namely, the $HOM(2)$ and the $SIM(2)$. The former is the so-called Homothety group, composed by the boost generator $K_{z}$ and the generators $T_{1}=K_{x}+J_{y}$, $T_{2}=K_{y}-J_{x}$, which form a group isomorphic to the group of translations in the plane \cite{ref1}. The latter, called the similitude group $SIM(2)$, is the $HOM(2)$ group improved by the addition of the $J_{z}$ generator. A relativity theory based on these two groups was  referred to by Cohen and Glashow as Very Special Relativity (VSR) \cite{ref1,ref2}.

Although built with the same Lorentz generators, the $SIM(2)$ is a 4-parameter group,
whereas the Lorentz group has six parameters. This reduction arises from the
generators $T_{1}$, $T_{2}$ that mix the boosts and angular momentum generators \cite{ref1}.
Sheikh-Jabbari and Tureanu argued that this crossing of the generators can arise due to a lightlike noncommutative structure in the space-time \cite{Jabbari}. The presence of the space-time translation is enough to guarantee the energy-momentum conservation and thereby the VSR yields to the desired conservation laws and classical test of the SR. The $SIM(2)$ generators preserve not only the velocity of light but also the null 4-vector $n_{\mu}=(1,0,0,1)$. Nevertheless, unlike the DSR, the dispersion relation is preserved. Thus, a spinless particle cannot perceive the Lorentz violating vector $n_{\mu}$ \cite{ref1}.

A key feature of the VSR is the absence of invariant tensors under $HOM(2)$ or $SIM(2)$
groups. In addition, the lack of the discrete symmetries leads to the violation of the celebrated CPT-theorem, which ensures a unitary and causal quantum field theory. Cohen-Glashow suggested including non-local operators to construct a unitary $SIM(2)$-invariant field theory. Indeed, the $T_{1,2}$ generators admit the ratios of the kind
$\frac{n \cdot p_{1}}{n \cdot p_{2}}$ as an invariant observable \cite{ref1}.
Therefore, a possible $SIM(2)$ invariant non-local operator is given by $N_{\mu}=\frac{n_{\mu}}{n \cdot\partial}$ that modifies the Dirac equation and can provide a non-vanishing mass to neutrino \cite{ref2}.

Nonetheless, in order to describe the natural phenomena, a field theory can not be composed of spinor fields only but of gauge fields as well. In this regard, a VSR extension of the Maxwell theory
was considered by Cheon et al. \cite{Cheon2009}.
The Maxwell equations and the gauge transformations are also modified by the inclusion of the non-local operators. Unlike the Lorentz-Maxwell theory, the augmented gauge symmetry allows the spin one particle to be massive, inasmuch as the VSR-parameter plays the mass role \cite{Cheon2009}. Lately, Alfaro and Rivelles addressed the non-Abelian
gauge fields that also acquire mass \cite{Alfaro2013}. By means of a $SIM(2)$ modified BRST transformation Alfaro-Rivelles found that the propagator and vertices have the same large momentum behavior as those of the usual Yang-Mills theory. The $SIM(2)$ non-Abelian gauge theory turned out to be asymptotically free and unitary as it should be \cite{Alfaro2013}.  Moreover, the framework of VSR has been explored in other contexts  covering different theoretical aspects: corrections to the  Thomas precession \cite{Alfaro2014}, curved spaces \cite{Gibbons:2007iu,Muck,Alvarez}, supersymmetry \cite{Cohen,Vohanka}, noncommutativity \cite{Jabbari,Das},  dark matter \cite{Ahluwalia:2010zn} and also in cosmology \cite{Kouretsis,Chang}.

Since the UV behavior (large momenta) of the $SIM(2)$ gauge theory agrees with the usual Lorentz-invariant theory, an interesting issue is to investigate for small deviations induced by VSR in the limit of low energies. Thus, we propose to study the low-energy regime of a $SIM(2)$ Dirac fermion in the presence of an external electromagnetic field, with the determination of the nonrelativistic (NR) Hamiltonian and the respective energy-level shifts. The foremost purpose of this letter is to analyze the low-energy effects able to distinguish VSR from the SR in the atomic scale. The NR Hamiltonian for electrons with spin-dependent masses and coupled to the electromagnetic fields was obtained by Fan et al. via calculation of the current matrix elements \cite{Fan}. However, the analysis of the hydrogen spectrum in the context of VSR was not addressed. We chose the energy transitions in the hydrogen atom due to its high precision which can also leads to stringent and new upper bounds upon the VSR-parameter. Besides, another worthy pursuit is to examine the effects of VSR on systems of condensed matter physics, an environment that has not been explored in this context yet.

This letter is organized as follows. In Section \ref{sec:theoretical-model}, we present
the $SIM(2)$-Dirac theory coupled with an external gauge field. Then, we attain the nonrelativistic limit of the VSR coupled Dirac equation and obtain the corresponding Hamiltonian. In Section \ref{sec3}, we carry out a perturbative analysis to find the first-order energy correction for the energy level of the hydrogen atom. In the last section \ref{sec:Conclusions} some final remarks and
perspectives are outlined.

\section{Dirac Spinor in Very Special Relativity\label{sec:theoretical-model}}

The $SIM(2)$-covariant Dirac equation for a free spin-$\nicefrac{1}{2}$
particle of mass $m$ is given by \cite{ref1,ref2}
\begin{equation}
\left[i\gamma^{\mu}\left(\partial_{\mu}+\frac{\lambda}{2}N_{\mu}\right)-m\right]\Psi(x)=0,\label{eq:Dirac1}
\end{equation}
where $N_{\mu}\equiv\frac{n_{\mu}}{n\cdot\partial}$ is a non-local differential operator with a
preferred null direction $n^{\mu}=(1,0,0,1)$, and the constant $\lambda$ measures
the strength of the VSR effects which it is ought to be very small \cite{ref1}.
In this letter we work out with the natural
unit system $(\hslash=c=1)$.

Taking into account that
$N\cdot N=0,$ $N\cdot\partial=1$, the Dirac spinor satisfies the Klein-Gordon
equation, with the shifted mass
\begin{equation}
\left[\partial_{\mu}\partial^{\mu}+M^{2}\right]\Psi(x)=0,\ \ \ \ \ \mbox{with}\ \ \ M^{2}=m^{2}+\lambda.\label{shiftedmass}
\end{equation}
Then, a massless particle can become massive due to the VSR symmetry what can be a solution to the neutrino mass problem \cite{ref2}.

Now we consider a charged Dirac particle in the presence of an
external electromagnetic field. As discussed in Refs. \cite{Cheon2009,Alfaro2013},
the $SIM(2)$-covariant action for the interacting Dirac field can be
written as
\begin{equation}
S_{\mbox{Dirac}}=\int d^{4}x\left[\bar{\Psi}(x)\left[i\gamma^{\mu}\left(D_{\mu}+\frac{\lambda}{2}\frac{n_{\mu}}{n\cdot D}\right)-m\right]\Psi(x)\right],\label{eq:DiracAction}
\end{equation}
where we have defined the covariant derivative by
\begin{equation}
D_{\mu}\Psi=\partial_{\mu}\Psi+ieA_{\mu}\Psi-ie\frac{\lambda}{2}n_{\mu}\left(\frac{1}{(n\cdot\partial)^{2}}n\cdot A\right)\Psi,
\end{equation}
such that under a $SIM(2)$-modified abelian gauge transformation\cite{Cheon2009,Alfaro2013}
\begin{equation}
A_{\mu}(x)\rightarrow A'_{\mu}(x)=A_{\mu}(x)+\partial_{\mu}\Lambda(x)+\frac{\lambda}{2}\left(N_{\mu}\Lambda\right)(x),
\end{equation}
$D_{\mu}\Psi$ and $\Psi$ transform according to
\begin{eqnarray}
D_{\mu}\Psi &   \rightarrow &   \left(D_{\mu}\Psi\right)'=e^{-ie\Lambda(x)}D_{\mu}\Psi,\\
\Psi(x) &   \rightarrow &   \Psi'(x)=e^{-ie\Lambda(x)}\Psi(x),
\end{eqnarray}
with $\Lambda(x)$ being an arbitrary function of x.

The Dirac equation corresponding to the action \eqref{eq:DiracAction}
is read as
\begin{equation}
\left[i\gamma^{\mu}\left(D_{\mu}+\frac{\lambda}{2}\frac{n_{\mu}}{n\cdot D}\right)-m\right]\Psi=0,\label{eq:DiracEq2}
\end{equation}
which it is our starting point to analyze the effects of VSR on the fermionic particle.

\subsection{Nonrelativistic limit}

In order to address the VSR modifications of the Dirac equation in an atomic system at the low-energy regime, we will determine the nonrelativistic Hamiltonian and find the corrections that it induces.

Writing the spinor $\Psi$ in terms of small $(\chi)$ and large $(\phi)$
two-component spinors,
\begin{equation}
\Psi=\begin{pmatrix}\phi\\
\chi
\end{pmatrix},
\end{equation}
and assuming the Dirac representation  of the $\gamma$-matrices, it is possible to show that the Eq. \eqref{eq:DiracEq2}
leads to two coupled equations for $\phi$ and $\chi$ in momentum space
\begin{align}
\left[E-eA_{0}-v_{0}-m\right]\phi-\left[\vec{\sigma}\cdot\left(\vec{p}-e\vec{A}-\vec{v}\right)\right]\chi & =0,\label{eq:Eqphi}\\
\left[\vec{\sigma}\cdot\left(\vec{p}-e\vec{A}-\vec{v}\right)\right]\phi-\left[E-eA_{0}-v_{0}+m\right]\chi & =0,\label{eq:Eqchi}
\end{align}
where we define
\begin{equation}
v_{0}\equiv\frac{\lambda}{2}\left[e\left(\frac{1}{\left(n\cdot p\right)^{2}}\left(n\cdot A\right)\right)+\frac{1}{n\cdot\left(p-eA\right)}\right],\label{parameter1-1}
\end{equation}

\begin{equation}
\vec{v}\equiv\frac{\lambda}{2}\left[e\left(\frac{1}{\left(n\cdot p\right)^{2}}\left(n\cdot A\right)\right)+\frac{1}{n\cdot\left(p-eA\right)}\right]\vec{n},\label{parameter2-1}
\end{equation}
 with $\vec{n}=(0,0,1)$ to represent the non-local VSR-differential
operators. We can represent all non-local terms through the following
identity:
\begin{equation}
\frac{1}{n\cdot\partial}=\int_{0}^{\infty}dae^{-an\cdot\partial}.
\end{equation}

In order to investigate the low-energy behavior of the system without
losing relativistic effects (like spin), we must access the nonrelativistic
limit, where the energy is given as $E=m+H$, with $H$ being the
nonrelativistic Hamiltonian. In the present case, where the rest mass
energy $m$ is much large than the other energies involved, we can
expand the non-local terms for $p_{z}\ll m$, $eA_{0}\ll m$, etc.,
and we obtain a local approximation for the VSR-differential operators
$v_{0}$ and $\vec{v}$, namely,
\begin{align}
v_{0} & \backsimeq\frac{\lambda e}{2m^{2}}\left(A_{0}-\vec{n}\cdot\vec{A}\right)+\frac{\lambda}{2m},\label{eq:vzero}\\
\vec{v} & \backsimeq\left[\frac{\lambda e}{2m^{2}}\left(A_{0}-\vec{n}\cdot\vec{A}\right)+\frac{\lambda}{2m}\right]\vec{n}.\label{eq:vvec}
\end{align}

From Eqs. \eqref{eq:Eqphi} and \eqref{eq:Eqchi}, the low-energy
condition leads to
\begin{equation}
\left(H-eA_{0}-v_{0}\right)\phi=\frac{1}{2m}\left(\vec{\sigma}\cdot\vec{\Pi}\right)\left(\vec{\sigma}\cdot\vec{\Pi}\right)\phi,
\end{equation}
where the canonical momentum is defined as $\vec{\Pi}\equiv\vec{p}-e\vec{A}-\vec{v}$.
After a short algebra, the nonrelativistic Hamiltonian for the particle
is reduced to the form,
\begin{align}
H & =\left\{ \left[\frac{1}{2m}\left(\vec{p}-e\vec{A}\right)^{2}-\frac{e}{2m}\vec{\sigma}\cdot\vec{B}+eA_{0}\right]\right.\nonumber \\
 & +\left.\frac{1}{2m}\vec{v}^{2}+v_{0}-\frac{1}{2m}\vec{p}\cdot\vec{v}-\frac{1}{m}\vec{v}\cdot\left(\vec{p}-e\vec{A}\right)-\frac{1}{2m}\vec{\sigma}\cdot\left(\nabla\times\vec{v}\right)\right\} .\label{eq:NonrelativisticH}
\end{align}
This is the full modified Hamiltonian, composed by the well-known
Pauli Hamiltonian (between brackets) corrected by the terms which
encode the effects due to the VSR where lies
our interest. It is worth to comment that a similar correction was
obtained in Ref. \cite{Manoel2006}, where the nonrelativistic Hamiltonian
was derived directly from the Dirac Lagrangian modified by the presence
of the term $v_{\mu}\bar{\psi}\gamma^{\mu}\psi$ in the framework
of the SME \cite{Colladay1997,Colladay1998}.
Nonetheless, this term does not contribute to any modification in the hydrogen spectrum. The Very Special Relativity, in its turn,
induces new effects. Note that the operators $v_{0}$ and $\vec{v}$,
defined by Eqs. \eqref{eq:vzero} and \eqref{eq:vvec}, represent
the components of a constant four-vector only in the free case. In the presence of an external electromagnetic field they become a
fixed background whose magnitude depends on the position of the particle.
Therefore, the VSR modifications of the energy levels of the hydrogen
atom is expected to be non-null.

\section{VSR-modification of energy levels\label{sec3}}

Once obtained the Hamiltonian for a charged particle under action
of an external electromagnetic field, we proceed to study the effects
of the new VSR terms on the energy levels of the hydrogen atom.

Although the VSR breaks the spherical symmetry, since the VSR effects are supposed to be small and we are interested in the local low-energy  level of the VSR Dirac equation, we consider the VSR interaction as a perturbation upon the usual spherically symmetric hydrogen atom under influence of the Colombian field.

In this regime, the unperturbed Colombian field $A^{\mu}=(\phi,\vec{0})$, with $\phi(r)=-\frac{e}{r}$,
leads to the local VSR-modified nonrelativistic Hamiltonian in the form
\begin{equation}
H=H_{0}+H_{\mbox{VSR}}+\mathcal{O}(\vec{v}^{2}),
\end{equation}
\begin{equation}
H_{0}=\frac{1}{2m_{e}}\vec{p}^{2}-\frac{e^{2}}{r},
\end{equation}
\begin{equation}
H_{\mbox{VSR}}=-\frac{1}{2m_{e}}\vec{p}\cdot\vec{v}-\frac{1}{m_{e}}\vec{v}\cdot\vec{p}-\frac{1}{2m_{e}}\vec{\sigma}\cdot\left(\nabla\times\vec{v}\right)+v_{0},
\end{equation}
with
\begin{equation}
v_{0}=\frac{\lambda}{2m_{e}}\left(1-\frac{e^{2}}{m_{e}}\frac{1}{r}\right),\ \ \ \ \vec{v}=v_{0}\hat{z}.
\end{equation}
It is worthwhile to note that, in this perturbative approach, the modifications produced by the VSR on the standard $1/r$ Coulomb potential of the proton are embodied in the Hamiltonian  $H_{\mbox{VSR}}$. Since in the framework of the Quantum Mechanics the Coulomb field is treated as an external field,
the VSR provides no dynamical correction to this field.

We calculate the corrections to the energy levels using
the eigenfunctions of $H_{0}$ and treat $H_{\mbox{VSR}}$ as a perturbation
in the context of the quantum mechanics. The nonrelativistic solutions
of hydrogen atom are well known and taking into account the presence
of the spin operator $\vec{\sigma}$ in the perturbation terms of
$H_{\mbox{VSR}}$, we can choose the complete set of commutating observables
$\{H_{0},L^{2},S^{2},L_{z},S_{z}\}$ and write the eigenfunctions
and the energy levels of $H_{0}$ as
\begin{equation}
H_{0}\Psi_{n,l,s,m_{l},m_{s}}=E_{n}\Psi_{n,l,s,m_{l},m_{s}},
\end{equation}
\begin{equation}
E_{n}=-\frac{e^{2}}{2n^{2}a_{0}},
\end{equation}
with $n=1,2,\ldots$, $l=0,1,\ldots,n-1,$ $s=\frac{1}{2}$ , $m_{l}=-l,-l+1,\ldots,l$
and $m_{s}=\pm\frac{1}{2}$ being the associated quantum numbers and
$a_{0}=1/e^{2}m_{e}$ is the Bohr radius. The eigenfunctions $\Psi_{n,l,s,m_{l},m_{s}}$
is explicitly written as
\begin{equation}
\Psi_{n,l,s,m_{l},m_{s}}=\psi_{n,l,m_{l}}(r,\theta,\phi)\chi_{sm_{s}},
\end{equation}
where the hydrogen 1-particle wave function is given by:
\begin{equation}
\psi_{n,l,m_{l}}(r,\theta,\phi)=R_{n,l}(r)Y_{l}^{m}(\theta,\phi),
\end{equation}
\begin{equation}
R_{n,l}(r)=\sqrt{\left(\frac{2}{na_{0}}\right)^{3}\frac{(n-l-1)!}{2n[(n+l)!]^{3}}}e^{-\frac{r}{na_{0}}}\left(\frac{2r}{na_{0}}\right)^{l}L_{n-l-1}^{2l+1}\left(\frac{2r}{ma_{0}}\right),
\end{equation}
with $L_{p}^{q}(x)$ are the associated Laguerre polynomials and $Y_{l}^{m}(\theta,\phi)$
are the spherical harmonics \cite{Gradshteyn}. Finally, the bispinors $\chi_{sm_{s}}$
can be represented by
\begin{eqnarray}
\chi_{s=\frac{1}{2},m_{s}=+\frac{1}{2}}=\left(\begin{array}{c}
1\\
0
\end{array}\right) & , & \chi_{s=\frac{1}{2},m_{s}=-\frac{1}{2}}=\left(\begin{array}{c}
0\\
1
\end{array}\right).
\end{eqnarray}

Since the energy of the hydrogen atom $E_{n}$ depends only of the
quantum number $n$, the unperturbed energy states are degenerate.
Considering the existence of the electron spin, the degeneration of the
energy levels is $2n^{2}$.

The purpose now is to investigate the possibility of lifts the degeneracy by the perturbation due to the
Hamiltonian $H_{\mbox{VSR}}$. To be specific, we
shall confine our attention to the corrections for the degenerate level $n=2$. At leading
order, it is achieved by calculating the eigenvalues of the $8\times8$
matrix $\Delta E_{\mbox{VSR}}$, which represents the correction $H_{\mbox{VSR}}$
inside the $2s$ and $2p$ subspaces (in spectroscopic notation).
Explicitly,

\begin{align}
\left(\Delta E_{\mbox{VSR}}\right)_{aa'} & =\frac{\lambda}{2m}\int d^{3}r\left\{ \Psi_{2,a}^{\dagger}(r,\theta\varphi)\left[\frac{ie^{2}}{2m^{2}}\frac{\cos\theta}{r^{2}}\right.\right.\nonumber \\
 & +\frac{i}{m}\left(1-\frac{e^{2}}{m}\frac{1}{r}\right)\left(\cos\theta\frac{\partial}{\partial r}-\frac{\sin\theta}{r}\frac{\partial}{\partial\theta}\right)\nonumber \\
 & +\frac{e^{2}}{2m^{2}}\left(\frac{\sin\theta\cos\varphi}{r^{2}}\sigma_{y}-\frac{\sin\theta\sin\varphi}{r^{2}}\sigma_{x}\right)\nonumber \\
 & +\left.\left.\left(1-\frac{e^{2}}{m}\frac{1}{r}\right)\right]\Psi_{2,a'}(r,\theta\varphi)\right\} ,\label{eq:DeltaE}
\end{align}
where $r$, $\theta$, $\varphi$ are spherical coordinates and the
label $a=(l,m_{l},m_{s})$ indicates the three degenerated quantum
numbers.

After a straightforward calculation, we find that only the second and fourth terms in \eqref{eq:DeltaE} yields to nonvanishing matrix
elements.
These terms represent the corrections $-\frac{1}{m}\vec{v}\cdot\vec{p}$
and $v_{0}$, respectively. For the nondiagonal elements, we have
\begin{equation}
\pm\frac{i\lambda e^{2}}{48a_{0}^{2}m_{e}^{3}},
\end{equation}
whereas the diagonal elements are equal to:
\begin{equation}
\frac{\lambda}{2m_{e}}\left(1-\frac{e^{2}}{4a_{0}m_{e}}\right).
\end{equation}

To complete our calculation, it is necessary diagonalize the matrix $\Delta E_{\mbox{VSR}}$ by solving the associated secular equation. It leads to the first-order energy correction and the zeroth-order
eigenstates, showed in Table \ref{eigenfunctions}.

\begin{table}
\begin{center}
\begin{tabular}{|ccc|}
\hline
Eigenstates &  & Corrections\tabularnewline
\hline
\hline
$\Psi_{n=2,l=1,m_{l}=\pm1,m_{s}=\pm\frac{1}{2}}$ & $\longleftrightarrow$ & $\frac{\lambda}{2m_{e}}\left(1-\frac{e^{2}}{4a_{0}m_{e}}\right)$\tabularnewline
\hline
$\bar{\Psi}_{+,-}=\frac{1}{\sqrt{2}}\left(\Psi_{n=2,l=1,m_{l}=0,m_{s}=\pm\frac{1}{2}}+i\Psi_{n=2,l=0,m_{l}=0,m_{s}=\pm\frac{1}{2}}\right)$ & $\longleftrightarrow$ & $\frac{\lambda}{2m_{e}}\left(1-\frac{e^{2}}{4a_{0}m_{e}}\right)-\frac{\lambda e^{2}}{48a_{0}^{2}m_{e}^{3}}$\tabularnewline
\hline
$\tilde{\Psi}_{+,-}=\frac{1}{\sqrt{2}}\left(\Psi_{n=2,l=1,m_{l}=0,m_{s}=\pm\frac{1}{2}}-i\Psi_{n=2,l=0,m_{l}=0,m_{s}=\pm\frac{1}{2}}\right)$ & $\longleftrightarrow$ & $\frac{\lambda}{2m_{e}}\left(1-\frac{e^{2}}{4a_{0}m_{e}}\right)+\frac{\lambda e^{2}}{48a_{0}^{2}m_{e}^{3}}$\tabularnewline
\hline
\end{tabular}
\par\end{center}
\caption{First-order corrected energies and their respective zero-order eigenfunctions.}
\label{eigenfunctions}
\end{table}

Therefore, the VSR-Hamiltonian terms turns the eightfold degeneracy
into a fourfold degeneracy of $\Psi_{2,1,m_{l}=\pm1,m_{s}=\pm\frac{1}{2}}$
states and two twofold degeneracies of states $\bar{\Psi}_{\pm}$
and $\tilde{\Psi}_{\pm}$. This result allows
us to conclude that the eightfold degeneracy of the levels $2s$
and $2p$ is partially removed by $H_{\mbox{VSR}}$, whereas the spin
degeneracy remains unaffected. The appearance of these energy corrections
is rather similar to the well-known Stark effect. This situation is
typical for the hydrogen atom considering two levels of opposite parities
and the same energy, such as the $2s$ and $2p$ levels.

Once the magnitude of these splittings depends directly on the $\lambda$
parameter, this theoretical outcome may be used to set up an upper
bound on the scale of VSR effects. According to the above results,
the $2p$ level, $m_{l}=1$ is shifted by a quantity equal to $\frac{\lambda}{2m_{e}}\left(1-\frac{e^{2}}{4a_{0}m_{e}}\right)$,
which is numerically $\simeq\lambda9.78\times10^{-7}(\mbox{eV})^{-1}$
(Bohr radius equal to $a_{0}\simeq0.0529nm\simeq2.69\times10^{-4}(\mbox{eV})^{-1}$).
Taking into account that spectroscopic experiments are able of measuring
shifts of $1$ part in $10^{10}$ in the spectrum, the energy correction
of Eq. \eqref{eq:DeltaE} will be undetectable if
\[
\left|\lambda9.78\times10^{-7}(\mbox{eV})^{-1}\right|<10^{-10}\mbox{eV},
\]
which implies an upper bound for the VSR-parameter
\begin{equation}
\label{bound}
\left|\lambda\right|<1.02\times10^{-4}(\mbox{eV})^{2}.
\end{equation}

The upper bound in Eq.\eqref{bound} is in a good agreement with that proposed by Cohen and Glashow based in measurements of the neutrino mass (see \cite{ref1} and references therein).  Indeed, the matching of these two distinct estimates suggests that some systems in condensed matter physics can be used to probe the VSR effects. Furthermore, it is still useful to compare the upper limit in Eq. \eqref{bound} with other values present in the literature. For example, Dunn and Mehen studied induced effects of VSR  in QED, as well as tree-level lepton family number violating interactions \cite{Mehen}. Their approach led to limiting values of $m^{2}_{\beta}/m^{2}_{e}<3.8\times10^{-12}$ (with $m_{\beta}$  being the electron based neutrino mass) from measurements of $g_{e}-2$ with trapped electrons, yield an upper bound $m^{2}_{\beta}<9.8\times10^{-1}(\mbox{eV})^{2}$, which is weaker than that obtained in \eqref{bound}.

\section{Conclusions and Perspectives\label{sec:Conclusions}}

In this work, we pursue the effects of the Very Special Relativity (VSR) on the Dirac field minimally coupled with a gauge field in the low-energy limit.
The main goal is to analyze how the VSR yields to the energy shift in the hydrogen atom under action of a Colombian electric field.

We start from the Dirac equation corresponding to the action of a charged particle in the presence of an external electromagnetic field. After that, in order to analyze the effects of VSR on the fermionic particle, we determine the correspondent nonrelativistic  modified Hamiltonian. Inspite of the lack of the spurions fields on the VSR, the local low-energy regime of the $SIM(2)$-Dirac equation exhibits fixed direction vector. It is important to mention that our setup, unlike the SME, give rise to new effects on the hydrogen spectrum. Since the VSR coupling constant should be small, the low-energy Hamiltonian can be split into an unperturbed term and a small perturbation. Then, by accomplishing perturbation methods we obtain the first-order corrections on the energy levels.

It turns out that the Hamiltonian term proportional to the spin coupling does not contribute to the energy shifts. On the other hand, the eightfold degeneracy of the second level is changed into a fourfold and two twofold degeneracies. This last result, not present in the SME approach, shows interesting physical implications of the VSR model. The present work also enable us to set up another upper bound on the VSR parameter $\lambda$ which unexpectedly fits well with a previous estimation done by
Cohen and Glashow \cite{ref1}. This achievement renders the atomic systems, besides the astrophysical dark matter \cite{Ahluwalia:2010zn}, as fair and interesting samples to probe VSR.

As a perspective we intend to investigate how the degeneracies behave when the non-local operators are included, a hallmark of VSR. Moreover, another worthy investigation is the analysis of the effects of the Deformed-VSR, whose symmetry group is given by the deformed $DSIM(2)$ \cite{Gibbons:2007iu}. Indeed, since the non-local effects in the Deformed-VSR are stronger than in usual VSR, and the degeneracies preserved in the present work could be broken.


\section*{Acknowledgments}
The authors thank the Coordena\c{c}\~ao de Aperfei\c{c}oamento de Pessoal de
N\'{i}vel Superior (CAPES), the Conselho Nacional de Desenvolvimento
Cient\'{i}fico e Tecnol\'ogico (CNPq), and Funda\c{c}\~ao Cearense de apoio ao Desenvolvimento Cient\'{\i}fico e Tecnol\'{o}gico (FUNCAP) for financial support.


\end{document}